\documentclass[titlepage,twoside,12pt]{article}
\usepackage{amssymb}
\usepackage{amsfonts}
\begin{document}

{\LARGE \bf Special Relativity in \\ \\ Reduced Power Algebras } \\ \\

{\bf Elem\'{e}r E ~Rosinger} \\ \\
{\small \it Department of Mathematics \\ and Applied Mathematics} \\
{\small \it University of Pretoria} \\
{\small \it Pretoria} \\
{\small \it 0002 South Africa} \\
{\small \it eerosinger@hotmail.com} \\ \\

{\bf Abstract} \\

Recently, [10,11], the Heisenberg Uncertainty relation and the No-Cloning property in Quantum Mechanics and Quantum
Computation, respectively, have been extended to versions of Quantum Mechanics and Quantum Computation which are
re-formulated using scalars in {\it reduced power algebras}, [2-9], instead of the usual real or complex scalars. Here,
the Lorentz coordinate transformations, fundamental in Special Relativity, are extended to versions of Special Relativity
that are similarly re-formulated in terms of scalars in reduced power algebras, instead of the usual real or complex
scalars. The interest in such re-formulations of basic theories of Physics are due to a number of important reasons,
[2-11]. Suffice to mention two of them : the difficult problem of so called "infinities in Physics" falls easily aside due
to the presence of infinitesimal and infinitely large scalars in such reduced power algebras, and the issue of fundamental
constants in physics, like Planck's $h$, or the velocity of light $c$, comes under a new focus which offers rather
surprising alternatives. \\ \\

{\bf 1. A Well Known Usual Deduction of the Lorentz \\
        \hspace*{0.5cm} Coordinate Transformations} \\

The following elementary way to obtain the Lorentz coordinate transformations is well known, [1]. Given two coordinate systems $S$ and $S\,'$ with
respective coordinates $( x, t )$ and $( x', t')$ in which the space $x$-axis and $x'$-axis are along the same line. We suppose that at time $t = t' =
0$ the origins $O$ and $O'$ of the two coordinate systems coincide, thus $x = x' = 0$. Let now $S'$ move along the$x$-axis in the positive direction with
the constant velocity $v$, and let two observers be respectively at $O$ and $O'$. \\

In that setup, at the initial moment $t = t' = 0$ and when $O$ and $O'$ coincide, a light signal is emitted from $O$. Its propagation within $S$ is then
given by \\

(1.1)~~~ $ x^2 = c^2  t^2 $ \\

where $c > 0$ is the velocity of light. \\

Now, in view of the Principle of Constancy of the Velocity of Light, in the coordinate system $S'$ the propagation of that light signal is according
to \\

(1.2)~~~ $ x'\,^2 =  c^2  t'\,^2 $ \\

Consequently, one must have \\

(1.3)~~~ $ x^2 - x'\,^2 = c^2 ( t^2 - t'\,^2 ) $ \\

However, at least for small values of $v$, when compared with $c$, we must have \\

(1.4)~~~ $ x' = k ( c, v ) ( x - v t ) $ \\

for some positive $k ( c, v ) \in \mathbb{R}$ that does not depend on $x, t, x', t'$, and which in addition is such that \\

(1.5)~~~ $ \lim_{v \to 0} k ( c, v ) = 1 $ \\

since (1.4) and (1.5) are implied by the respective non-relativistic Galilean coordinate transformation.

Now in view of the Principle of Relativity of Motion, we can suppose that $S'$ is fixed, and $S$ is moving along the $x'$-axis and in the negative
direction, with velocity $-v$. In that case, similar with (1.4), we obtain \\

(1.6)~~~ $ x = k ( c, v ) ( x' + v t' ) $ \\

By squaring (1.4) and (1.6), we obtain \\

(1.7)~~~ $ x'\,^2 + k ( c, v )^2 x^2 - 2 k ( c, v ) x x' = k ( c, v )^2 v^2 t^2 $ \\

(1.8)~~~ $ x^2 + k ( c, v )^2 x'\,^2 - 2 k ( c, v ) x x' = k ( c, v )^2 v^2 t'\,^2 $ \\

thus by subtracting the (1.7) from (1.8), it follows that \\

(1.9)~~~ $ ( x^2 - x'\,^2 ) ( k ( c, v )^2 - 1 ) = k ( c, v )^2 v^2 ( t^2 - t'\,^2 ) $ \\

and then in view of (1.3), we obtain \\

(1.10)~~~ $ c^2 ( k ( c, v )^2 - 1 ) =  k ( c, v )^2 v^2  $ \\

or \\

(1.11)~~~ $ ( c^2 - v^2 ) k ( c, v )^2 = c^2 $ \\

In this way \\

(1.12)~~~ $ k ( c, v ) = c / ( c^2 - v^2 )^{1/2} = 1 / ( 1 - v^2 / c^2 )^{1/2} $ \\

which obviously satisfies (1.5). \\

The {\it space coordinate} Lorentz transformation results now from (1.4) and (1.12), namely \\

(1.13)~~~ $ x' = ( x - v t ) / ( 1 - v^2 / c^2 )^{1/2} $ \\

In order to obtain the {\it time coordinate} Lorentz transformation, it will be convenient to proceed in full algebraic detail, and with a special
attention to the operations of {\it division} and {\it square root} involved. For that purpose, we replace $x'$ in (1.6) with its value from (1.4). The
result is \\

(1.14)~~~ $ \begin{array}{l}
                               x =  k ( c, v ) ( k ( c, v ) ( x - v t ) + v t' ) = \\ \\
                               ~~~~~~~~~ = k ( c, v )^2 x - k ( c, v )^2 v t + k ( c, v ) v t'
            \end{array} $ \\

or \\

(1.15)~~~ $ k ( c, v ) v t' = k ( c, v )^2 v t - ( k ( c, v )^2 - 1 ) x $ \\

Thus dividing by $k ( c, v ) v$, one has \\

(1.16)~~~ $ t' = k ( c, v ) t - (  k ( c, v )^2 - 1 ) / ( k ( c, v ) v ) ) x $ \\

Dividing in (1.11) by $c^2 - v^2$, results that \\

(1.17)~~~ $ k ( c, v )^2 = c^2 / ( c^2 - v^2 ) $ \\

and then \\

(1.18)~~~ $  k ( c, v )^2 - 1 = v^2 / ( c^2 - v^2 ) $ \\

Now (1.16), (1.12) yield the desired {\it time coordinate} Lorentz transformation \\

(1.19)~~~ $ t' = ( t - v x / c^2 ) / ( 1 - v^2 / c^2 )^{1/2} $ \\ \\

{\bf 2. Extending the Lorentz Coordinate Transformations to \\
        \hspace*{0.5cm} Reduced Power Algebras} \\

Let us consider instead of the field $\mathbb{R}$ of usual real numbers an arbitrary reduced power algebra $\mathbb{A}_{\cal F}$, see (A.1.4) in the
Appendix. In other words, we shall model both space and time with such algebras $\mathbb{A}_{\cal F}$, instead of modelling them with the field
$\mathbb{R}$ of usual real numbers. \\
Here it is important to note that, in general, such algebras $\mathbb{A}_{\cal F}$ need {\it not} be linearly or totally ordered, see (A.4.1) - (A.4.4)
in the Appendix. Furthermore, when they are not linearly or totally ordered, that is, when the respective filters ${\cal F}$ are not ultrafilters, then
the corresponding algebras $\mathbb{A}_{\cal F}$ need {\it not} be one dimensional vector spaces, as is of course the case of $\mathbb{R}$. \\
It follows that the extension of the Lorentz coordinate transformations to reduced power algebras opens up a rather wide realm, one in which time, as
much as each individual coordinate, may be {\it multi-dimensional}, and in fact, even {\it infinite dimensional}. \\
Speculations regarding the possible meaning of such considerable extensions can, therefore, be diverse and rather numerous. One of them, coming from the
multi-dimensionality of time, may be that it could possibly model {\it parallel universes} ... \\

And now, let us return to the aimed extension the Lorentz coordinate transformations to arbitrary reduced power algebras. \\

In this regard, it is sufficient to note that all the algebraic operations in section 1 above, operations leading to the usual Lorentz coordinate
transformations in (1.13), (1.19), can automatically be replicated in all the reduced power algebras $\mathbb{A}_{\cal F}$, except when divisions and
square roots are involved. Indeed, when divisions are involved in these algebras one has to consider the presence in them of {\it zero divisors} and
{\it non-invertible} elements, see section A.2. in the Appendix. As for square roots, one has to proceed according to section A.5. in the Appendix. \\ \\

{\bf 3. Comments} \\

{\bf 3.1. Why Hold to the Archimedean Axiom ?} \\

It is seldom realized, especially among physicists, that ever since ancient Egypt and the axiomatization of Geometry by Euclid, we keep holding to the
Archimedean Axiom. This axiom, in simplest terms, such as of a partially ordered group $G$, for instance, means the following property \\

(3.1.1)~~~ $ \exists~~ u \in G,~ u \geq 0 ~:~ \forall~~ x \in G ~:~ \exists~~ n \in \mathbb{N} ~:~ x \leq n u $ \\

or in other words, there exists a "path length" $u$, so that every element $x$ in the group can be "overtaken" by a finite number $n$ of "steps" of
"length" $u$. Clearly, if $G$ is the set $\mathbb{R}$ of usual real numbers considered with the usual addition, then one can take as $u$ any positive
number. \\
As is known, Geometry in ancient Egypt was important in connection with the yearly flood of the Nile and the subsequent need to redraw the boundaries of
agricultural land. And for such a purpose, the Archimedean Axiom is obviously useful. \\

The question, however, is :

\begin{quote}

Why hold to that axiom when dealing with such modern and highly non-intuitive theories of Physics, as Special and General
Relativity, or Quantum Mechanics and Quantum Field Theory ? \\
Is there any physical type reason in such modern theories for holding to the Archimedean Axiom ?

\end{quote}

Indeed, one of the inevitable consequences of the Archimedean Axiom is that "infinity" is not a usual scalar, be it real or complex. Thus all usual
algebraic and other operations do rather as a rule break down when reaching "infinity". And this elementary and inevitable fact leads to the long
festering problem of the so called "infinities in Physics", a problem which is attempted to be dealt with by various "re-normalization" methods, or by
what is an exceedingly complex and so far not yet successful venture, namely, String Theory. \\

On the other hand, the moment one simply frees oneself from the Archimedean Axiom, and starts to deal with scalars such as those given by various reduced
power algebras, the mentioned troubles with "infinity" disappear. Indeed, since the Archimedean Axiom is no longer present in such algebras, these
algebras have a rich structure of "infinitesimals" and "infinitely large" scalars, all of which are subjected to the usual algebraic and other
operations, just as if they were usual real or complex numbers. \\

{\bf 3.2. Two Alternatives When Freed From the Archimedean \\
          \hspace*{0.8cm} Axiom} \\

The above way the Lorentz Coordinate Transformations have been extended to space-times built upon scalars given by reduced
power algebras may at first seem to be both trivial and without interest. And the same appearance may arise with the
extension to such space-times of the Heisenberg Uncertainty and No-Cloning, in [10], respectively, [11]. \\
Here however, one should note the following. \\

First, even the multiplication in such reduced power algebras is no longer trivial. Indeed, such algebras can have zero
divisors, see section A.2. in the Appendix. Consequently, it may easily happen that, although $c, v, x, t, k ( c, v )
\neq 0$, we will nevertheless have the products in which such quantities appear, and the respective products vanish, contrary to what happens in the usual case
when scalars given by real numbers are employed. And clearly, such a vanishing of certain products may invalidate
subsequent formulas, or at best, give them a different meaning from the usual one. \\
Also, mathematical expressions in various theories of Physics can contain operations other than mere multiplication, and such operations
can have new properties and meanings, when performed in reduced power algebras. \\

Therefore, here, we may obviously face two rather different alternatives, namely

\begin{itemize}

\item the new properties and meanings in reduced power algebras do not correspond to any possible physical meaning,
    
\end{itemize}

or on the contrary

\begin{itemize}

\item such new properties and meanings which appear in reduced power algebras may possibly correspond to not yet explored
physical realities.

\end{itemize}

We shall in the sequel mention several possible such new physical interpretations, if not in fact, possible 
realities. \\ \\

{\bf 3.3. Increased and Decreased Precision in Measurements} \\

As a general issue, relating not only to Relativity or the Quanta, the presence of infinitesimal and infinitely large
scalars in reduced power algebras may correspond to a new possibility of having no less than {\it two} radically different
kind of measurements when it comes to their {\it relative precision}. \\

Namely, one has an {\it increased precision} in measurement, when measurement is done in terms of usual finite scalars,
and one obtains as result some infinitesimal scalar in such algebras. \\

Alternatively, the presence of infinitely large scalars in such algebras may simply indicate that they were obtained in
terms of finite scalars, and thus are but the result of a measurement with {\it decreased precision}. \\

In this regard, we can therefore have the following {\it relative} situations

\begin{itemize}

\item infinitesimal scalars are the result of increased precision measurements done in terms of finite or infinite scalars,

\item finite scalars are the result of increased precision measurements done in terms of infinite scalars,

\item finite or infinitely large scalars are the result of decreased precision measurements done in terms of infinitesimal
scalars,

\item infinitely large scalars are the result of decreased precision measurements done in terms of infinitesimal or finite
scalars.

\end{itemize}

and surprisingly, one can also have the following {\it relative} situations

\begin{itemize}

\item infinitesimal scalars are the result of increased precision measurements done in terms of some less infinitesimal
scalars,

\item infinitesimal scalars are the result of decreased precision measurements done in terms of some more infinitesimal
scalars,

\item infinitely large scalars are the result of increased precision measurements done in terms of some more infinitely
large scalars,

\item infinitely large scalars are the result of decreased precision measurements done in terms of some less infinitely
large scalars.

\end{itemize}

Indeed, one of the basic features of reduced power algebras is precisely their complicated and rich {\it self-similar}
structure which distinguishes not only between infinitesimal, finite and infinitely large scalars, but also within the
infinitely small scalars themselves, and similarly, within the infinitely large scalars. Specifically,
infinitesimal scalars can be infinitely smaller, or on the contrary, infinitely larger than other infinitesimals. And
similarly, infinitely large scalars can be infinitely smaller, or on the contrary, infinitely larger than other infinitely
large scalars. \\

Here, however, we can note that such a possible interpretation of increased, or decreased precision which is {\it
relative}, is in fact not new. Indeed, in terms of usual scalars, be they real or complex, there is a marked dichotomy
between finite scalars, and on the other hand, the so called "infinities" which may on occasion arise from operations with
finite scalars. And such simple "formulas" like $\infty + 1 = \infty$, are in fact expressing that fact. Namely, on one
hand, from the point of view of "infinity", the finite number $1$ has such an increased precision as to be irrelevant with
respect to addition, while on the other hand, from the point of view of the finite number $1$, the "infinity" has such a
decreased precision as to alter completely the result when involved in addition. \\ \\

{\bf 3.4. The Issue of Universal Constants} \\

Given the above possibilities in interpretation leading to relative precision measurement - be it as such an increased or a decreased one - one
can reconsider the status of certain universal physical constants, such as for instance, the Planck constant $h$ and the
constant $c$ giving the velocity of light in vacuum. \\
Indeed, when considered from our everyday macroscopic experience, $h$ is supposed to be unusually small, while on the
contrary, $c$ is very large. Consequently, one may see $h$ as a sort of "infinitesimal", while $c$ then looks like
"infinitely large". \\
The fact is that, within reduced power algebras, such an alternative view of $h$ and $c$ is possible. Therefore, one may
find it appropriate to explore the possible physical meaning, or otherwise, that may possibly be associated with such an
interpretation. \\ \\

{\bf Appendix : Zero Divisors, Units and other Properties \\
                \hspace*{2.5cm} in Reduced Power Algebras} \\

{\bf A.1. Construction of Reduced Power Algebras} \\

The general construction of {\it reduced power algebras} goes as follows, [2-11]. Let $\Lambda$ be any {\it infinite} set. Let ${\cal F}$ be any filter on
$\Lambda$, such that \\

(A.1.1)~~~ $ {\cal F}_{re} ( \Lambda ) \subseteq {\cal F} $ \\

where \\

(A.1.2)~~~ $ {\cal F}_{re} ( \Lambda ) = \{~ I \subseteq \Lambda ~~|~~ \Lambda \setminus I ~~\mbox{is finite} ~\} $ \\

is called the Frech\`{e}t filter on $\Lambda$. \\

We define on $\mathbb{R}^\Lambda$ the corresponding equivalence relation $\approx_{\cal F}$ by \\

(A.1.3)~~~ $ x \approx_{\cal U} y ~~\Longleftrightarrow~~
                        \{~ \lambda \in \Lambda ~|~ x ( \lambda ) = y ( \lambda ) ~\} \in {\cal F} $ \\

where $x, y \in \mathbb{R}^\Lambda$. \\

Then, through the usual quotient construction, we obtain the {\it reduced power algebra} \\

(A.1.4)~~~ $ \mathbb{A}_{\cal F} = \mathbb{R}^\Lambda / \approx_{\cal F} $ \\

which has the following two properties. \\

The mapping \\

(A.1.5)~~~ $ \mathbb{R} \ni r \longmapsto ( u_r )_{\cal F} \in \mathbb{A}_{\cal F} $ \\

is an {\it embedding of algebras} in which $\mathbb{R}$ is a strict subset of $\mathbb{A}_{\cal F}$, where $u_r \in
\mathbb{R}^\Lambda$ is defined by $u_r ( \lambda ) = r$, for $\lambda \in \Lambda$, while $( u_r )_{\cal F}$ is the coset
of $u_r$ with respect to the equivalence relation $\approx_{\cal F}$. \\

Further, on $\mathbb{A}_{\cal F}$ we have the {\it partial order} which is compatible with the algebra structure, namely \\

(A.1.6)~~~ $ ( x )_{\cal F} \leq ( y )_{\cal F} ~~\Longleftrightarrow~~
             \{~ \lambda \in \Lambda ~|~ x ( \lambda ) \leq y ( \lambda ) ~\} \in {\cal F} $ \\

where $x, y \in \mathbb{R}^\Lambda$. \\

As is well known \\

(A.1.7)~~~ $ \mathbb{A}_{\cal F} ~~\mbox{is a field} ~~~\Longleftrightarrow~~~ {\cal F} ~~\mbox{is an ultrafilter on}~~ \Lambda $ \\

consequently \\

(A.1.8)~~~ $ \mathbb{A}_{\cal F} ~~\mbox{has zero divisors} ~~~\Longleftrightarrow~~~ {\cal F} ~~\mbox{is not an ultrafilter on}~~ \Lambda $ \\

It will be useful to consider the {\it non-negative} elements in $\mathbb{A}_{\cal F}$, given by \\

(A.1.9)~~~ $ \mathbb{A}^+_{\cal F} =
                    \{~ ( x )_{\cal F} ~|~ x \in \mathbb{R},~~~ \{~ \lambda \in \Lambda ~|~ x ( \lambda ) \geq 0 ~\} \in {\cal F} ~\} $ \\ \\

{\bf A.2. Zero Divisors and Units in $\mathbb{A}_{\cal F}$} \\

Let ${\cal F}$ be a filter on $\Lambda$ which satisfies (A.1.1) and is not an ultrafilter on $\Lambda$. Given any $x \in \mathbb{R}^\Lambda$, we denote \\

(A.2.1)~~~ $ Z ( x ) = \{~ \lambda \in \Lambda ~~|~~ x ( \lambda ) = 0 ~\} \subseteq \Lambda $ \\

and obviously, we have the following four alternatives \\

(A.2.2.1)~~~ $ Z ( x ) \in {\cal F} $ \\

(A.2.2.2)~~~ $ Z ( x ) \notin {\cal F} $ \\

(A.2.2.3)~~~ $ \Lambda \setminus Z ( x ) \in {\cal F} $ \\

(A.2.2.4)~~~ $ \Lambda \setminus Z ( x ) \notin {\cal F} $ \\

Since ${\cal F}$ is not an ultrafilter, alternatives (A.2.2.1) and (A.2.2.3) are not incompatible. Therefore, the same applies to alternatives (A.2.2.2) and
(A.2.2.4). It follows that we have the mutually exclusive four alternatives \\

(A.2.3.1)~~~ $ Z ( x ) \in {\cal F} ~~\mbox{and}~~ \Lambda \setminus Z ( x ) \in {\cal F} $ \\

(A.2.3.2)~~~ $ Z ( x ) \in {\cal F} ~~\mbox{and}~~ \Lambda \setminus Z ( x ) \notin {\cal F} $ \\

(A.2.3.3)~~~ $ Z ( x ) \notin {\cal F} ~~\mbox{and}~~ \Lambda \setminus Z ( x ) \in {\cal F} $ \\

(A.2.3.4)~~~ $ Z ( x ) \notin {\cal F} ~~\mbox{and}~~ \Lambda \setminus Z ( x ) \notin {\cal F} $ \\

Now in view of (A.1.3), we have \\

(A.2.4)~~~ $ Z ( x ) \in {\cal F} ~~~\Longleftrightarrow~~~ ( x )_{\cal F} = 0 \in \mathbb{A}_{\cal F} $ \\

thus alternatives (A.2.3.1) and (A.2.3.2) are clarified in their consequence. \\

Let us now consider (A.2.3.3) and define $y \in \mathbb{R}^\Lambda$ by \\

(A.2.5)~~~ $ y ( \lambda ) ~=~ \begin{array}{|l}
                                     ~ 1 / x ( \lambda ) ~~\mbox{if}~~ \lambda \in \Lambda \setminus Z ( x ) \\ \\
                                     ~\mbox{arbitrary otherwise}
                            \end{array} $ \\

then (A.1.3), (A.2.4) give \\

(A.2.6)~~~ $ ( x )_{\cal F},~ ( y )_{\cal F} \neq 0 \in \mathbb{A}_{\cal F},~~
                                                    ( x )_{\cal F} \, ( y )_{\cal F} = 1 \in \mathbb{A}_{\cal F} $ \\

thus  $( x )_{\cal F}$ is an {\it invertible element}, or a {\it unit} in $\mathbb{A}_{\cal F}$, and  $( ( x )_{\cal F} )^{-1} = ( y )_{\cal F}$ . \\

In the case of (A.2.3.4), let us define $y \in \mathbb{R}^\Lambda$ by \\

(A.2.7)~~~ $ y ( \lambda ) ~=~ \begin{array}{|l}
                                     ~ 0 ~~\mbox{if}~~ \lambda \in \Lambda \setminus Z ( x ) \\ \\
                                     ~ 1 ~~\mbox{if}~~ \lambda \in Z ( x )
                            \end{array} $ \\

then (A.1.3), (A.2.4) give \\

(A.2.8)~~~ $ ( x )_{\cal F},~ ( y )_{\cal F} \neq 0 \in \mathbb{A}_{\cal F},~~
                                                    ( x )_{\cal F} \, ( y )_{\cal F} = 0 \in \mathbb{A}_{\cal F} $ \\

thus  $( x )_{\cal F}$ is a {\it zero divisor} in $\mathbb{A}_{\cal F}$. \\

It follows that the set of {\it units}, or {\it invertible elements} in $\mathbb{A}_{\cal F}$ is given by \\

(A.2.9)~~~ $ \mathbb{A}^u_{\cal F} = \{~ ( x )_{\cal F} ~|~ x \in \mathbb{R}^\Lambda,~
                 Z ( x ) \notin {\cal F},~ \Lambda \setminus Z ( x ) \in {\cal F} ~\} $ \\

while the set of {\it zero divisors} in $\mathbb{A}_{\cal F}$ is given by \\

(A.2.10)~~~ $ \mathbb{A}^{zd}_{\cal F} = \{~ ( x )_{\cal F} ~|~ x \in \mathbb{R}^\Lambda,~
                 Z ( x ) \notin {\cal F},~ \Lambda \setminus Z ( x ) \notin {\cal F} ~\} $ \\

and clearly, we have the following partition in three disjoint subsets \\

(A.2.11)~~~ $ \mathbb{A}_{\cal F} = \{ 0 \} \,\bigcup\, \mathbb{A}^{zd}_{\cal F} \,\bigcup\, \mathbb{A}^u_{\cal F} $ \\ \\

{\bf A.3. Infinitesimals and Infinitely Large Scalars} \\

The reduced power algebras $\mathbb{A}_{\cal F}$ contain strictly as a subfield the field $\mathbb{R}$ of usual real numbers. In addition, the reduced
power algebras $\mathbb{A}_{\cal F}$ contain vast amounts of {\it infinitesimal}, as well as {\it infinitely large} scalars. \\

In case in (A.1.3), and in the sequel, we replace $\mathbb{R}$ with $\mathbb{C}$, and thus $\mathbb{C}^\Lambda$ takes the place of $\mathbb{R}^\Lambda$, then we obtain reduced power algebras which contain strictly the field
$\mathbb{C}$ of usual complex numbers. And again, the reduced power algebras will contain vast amounts of {\it infinitesimal}, as well as {\it
infinitely large} scalars. \\ \\

{\bf A.4. Reduced Power Fields} \\

The following properties are equivalent : \\

(A.4.1)~~~ $ {\cal F} $ is an {\it ultrafilter} on $ \Lambda $ \\

(A.4.2)~~~ $ \mathbb{A}^{zd}_{\cal F} = \phi,~~~ \mathbb{A}_{\cal F} = \{ 0 \} \,\bigcup\, \mathbb{A}^u_{\cal F} $~ is a field \\

(A.4.3)~~~ For every ~$ x \in \mathbb{R}^\Lambda $,~ the four alternatives (A.2.3.1) - (A.2.3.4) reduce to the following two, namely, (A.2.3.2),
(A.2.3.3), that is : \\

~~~~~~~~~ $ Z ( x ) \in {\cal F} ~~\mbox{and}~~ \Lambda \setminus Z ( x ) \notin {\cal F} $ \\

~~~~~~~~~ $ Z ( x ) \notin {\cal F} ~~\mbox{and}~~ \Lambda \setminus Z ( x ) \in {\cal F} $ \\

(A.4.4)~~~ The partial order $\leq_{\cal F}$ in (A.1.6) is a linear, or total order on the reduced power field $ \mathbb{A}_{\cal F} $ \\ \\

{\bf A.5. Exponential Functions} \\

In view of (A.1.9), one can obviously define the exponentiation \\

(A.5.1)~~~ $ \mathbb{A}^+_{\cal F} \times \mathbb{A}^+_{\cal F} \ni ( ( x )_{\cal F}, ( y )_{\cal F} )
                                              \longmapsto ( z )_{\cal F} = ( ( x )_{\cal F} )^{( ( y )_{\cal F} )}\in \mathbb{A}^+_{\cal F} $ \\

by \\

(A.5.2)~~~ $ z ( \lambda ) = ( x ( \lambda ) )^{( y ( \lambda ) )},~~~ \lambda \in I $ \\

where $I \in {\cal F}$ is such that \\

(A.5.3)~~~ $  x ( \lambda ),~ y ( \lambda ) \geq 0,~~~ \lambda \in I $ \\ \\

\end{document}